\documentclass[11pt]{article}

\usepackage{etex}
\usepackage{graphicx}
\usepackage{amsmath,amsthm}
\usepackage{amsfonts}
\usepackage{amssymb}
\usepackage{array}
\usepackage{multirow}
\usepackage{longtable}
\usepackage{setspace}
\usepackage{subfigure}
\usepackage{pgf,tikz}
\usepackage{mathrsfs}
\usetikzlibrary{arrows}
\usetikzlibrary{calc}
\usepackage[rflt]{floatflt}
\usepackage{color, colortbl}
\usepackage{fullpage}
\usepackage[final]{pdfpages}
\usepackage[algo2e,ruled,vlined]{algorithm2e}
\usepackage{hyperref}

\usepackage{verbatim}
\usepackage{enumerate}

\theoremstyle{definition}
\newtheorem{defn}{Definition}

\usepackage[mathlines]{lineno}
\theoremstyle{plain}
\newtheorem{thm}{Theorem}

\newtheorem{ob}[thm]{Observation}
\newtheorem{prop}[thm]{Proposition}

\newtheorem{cor}[thm]{Corollary}
\newtheorem{prob}{Problem}

\newcommand{\sub}{\mathrm{sub}}

\definecolor{red}{rgb}{1,0,0}

\begin{document}

\title{ The sub-$k$-domination number of a graph with \\ applications to $k$-domination}
\author{
David Amos \thanks{Department of Mathematics, Texas A\&M University, USA. \texttt{(dave.amos@live.com)}}
\and 
John Asplund\thanks{Department of Technology and Mathematics,
Dalton State College, USA.
\texttt{(jasplund@daltonstate.edu)}}
\and
Boris Brimkov \thanks{Department of Computational and Applied Mathematics, Rice University, USA. \texttt{(boris.brimkov@rice.edu)}} \and 
Randy Davila \thanks{Department of Pure and Applied Mathematics, University of Johannesburg, South Africa.}$\;\,^,$
\thanks{Department of Mathematics, Texas State University, USA.
\texttt{(rrd32@txstate.edu)}}
}
\date{}

\maketitle

\begin{abstract}

In this paper we introduce and study a new graph invariant derived from the degree sequence of a graph $G$, called the \emph{sub-$k$-domination number} and denoted $\sub_k(G)$. We show that $\sub_k(G)$ is a computationally efficient sharp lower bound on the $k$-domination number of $G$, and improves on several known lower bounds. We also characterize the sub-$k$-domination numbers of several families of graphs, provide structural results on sub-$k$-domination, and explore properties of graphs which are $\sub_k(G)$-critical with respect to addition and deletion of vertices and edges. 
\end{abstract}

{\small \textbf{Keywords:} sub-$k$-domination number, $k$-domination number, degree sequence index strategy}\\
\indent {\small \textbf{AMS subject classification: 05C69}}

\section{Introduction}
Domination is one of the most well-studied and widely applied concepts in graph theory. A set $S\subseteq V(G)$ is \emph{dominating} for a graph $G$ if every vertex of $G$ is either in $S$, or is adjacent to a vertex in $S$. A related parameter of interest is the \emph{domination number}, denoted $\gamma(G)$, which is the cardinality of the smallest dominating set of $G$. Much of the literature on domination is surveyed in the two monographs of Haynes, Hedetniemi, and Slater \cite{DomBook1, DomBook2}. For more recent results on domination, see \cite{Dorfling,Du,Glebov,Stevanovic} and the references therein.

In 1984, Fink and Jacobson \cite{FinkJacobson} generalized domination by introducing the notion of $k$-domination and its associated graph invariant, the $k$-domination number. Given a positive integer $k$, $S\subseteq V(G)$ is a \emph{$k$-dominating set} for a graph $G$ if every vertex not in $S$ is adjacent to at least $k$ vertices in $S$. The minimum cardinality of a $k$-dominating set of $G$ is the \emph{$k$-domination number} of $G$, denoted $\gamma_k(G)$. When $k=1$, the $1$-domination number is precisely the domination number; that is, $\gamma_1(G)=\gamma(G)$. Like domination, $k$-domination has also been extensively studied; for results on $k$-domination related to this paper, we refer the reader to \cite{CaroRoditty, Delavina, Favaron, Adriana, Pepper, Rautenbach}.

Computing the $k$-domination number is $NP$-hard \cite{NP}, and as such, many researchers have sought computationally efficient upper and lower bounds for this parameter. In general, the degree sequence of a graph can be a useful tool for bounding $NP$-hard graph invariants. For example, the \emph{residue} and \emph{annihilation number} of a graph are derived from its degree sequence, and are respectively lower and upper bounds on the graph's independence number (cf. \cite{FavaronResidue,PepperThesis}). Another example is a lower bound on the domination number due to Slater \cite{Slater}, which will be discussed in the sequel. Recently, Caro and Pepper \cite{DSI} introduced the \emph{degree sequence index strategy}, or DSI-strategy, which provides a unified framework for using the degree sequence of a graph to bound $NP$-hard invariants. In this paper we introduce a new degree sequence invariant called the sub-$k$-domination number, which is a sharp lower bound on the $k$-domination number; our investigation contributes to the known literature on both degree sequence invariants and domination. 

Throughout this paper all graphs are simple and finite. Let $G=(V(G),E(G))$ be graph. Two vertices $v$ and $w$ in $G$ are \emph{adjacent}, or \emph{neighbors}, if there exists an edge $vw\in E$. A vertex is an \emph{isolate} if it has no neighbors. The \emph{complement} of $G$ is the graph $\overline{G}$ with the same vertex set, in which two vertices are adjacent if and only if they are not adjacent in $G$. A set $S\subseteq V(G)$ is \emph{independent} if no two vertices in $S$ are adjacent; the cardinality of the largest independent set in $G$ is denoted $\alpha(G)$. For any edge $e\in E(G)$, $G-e$ denotes the graph $G$ with the edge $e$ removed; For any vertex $v\in V(G)$, $G-v$ denotes the graph $G$ with the vertex $v$ and all edges incident to $v$ removed; for any edge $e\in E(\overline{G})$, $G+e$ denotes the graph $G$ with the edge $e$ added. The \emph{degree} of a vertex $v$, denoted $d(v)$, is the number of vertices adjacent to $v$. We will use the notation $n(G)=|V(G)|$ to denote the order of $G$, $\Delta(G)$ to denote the maximum degree of $G$, and $\delta(G)$ to denote the minimum degree of $G$; when there is no scope for confusion, the dependence on $G$ will be omitted. We will also use $d_i$ to denote the $i^\mathrm{th}$ element in the \emph{degree sequence} of $G$, denoted $D(G) = \{\Delta = d_1\geq d_2\geq \cdots \geq d_n = \delta\}$, which lists the vertex degrees in non-increasing order. We may abbreviate $D(G)$ by only writing distinct degrees, with the number of vertices realizing each degree in superscript. For example, the star $K_{n-1,1}$ may have its degree sequence written as $D(K_{n-1,1}) = \{n-1, 1^{n-1}\}$, and the complete graph $K_n$ may have degree sequence written as $D(K_n) = \{ (n-1)^{n}\}$. For other graph terminology and notation, we will generally follow \cite{MHAYbookTD}.

This paper is organized as follows. In the next section, we introduce the sub-$k$-domination number of a graph and show that it is a lower bound on the $k$-domination number. In Section 3, we characterize the sub-$k$-domination numbers of several families of graphs and provide other structural results on sub-$k$-domination. In Section 4, we compare the sub-$k$-domination number to other known lower bounds on the $k$-domination number. In Section 5, we explore the properties of $\sub_k(G)$-critical graphs. We conclude with some final remarks and open questions in Section 6.

%%%%%%%%%%%%%%%

\section{Sub-$k$-domination}
In this section we introduce the sub-$k$-domination number of a graph and prove that it is a lower bound on the $k$-domination number. We first recall a definition and result due to Slater \cite{Slater}, which is a special case of our result. For consistency in terminology, we will refer to Slater's definition as the \emph{sub-domination number} of a graph; this invariant was originally denoted $sl(G)$, and for our purposes will be denoted $\sub(G)$.

\begin{defn}[{\normalfont\cite{Slater}}]\label{SlaterDefn}
The \emph{sub-domination number} of a graph $G$ is defined as
\begin{equation*}
\sub(G) = \min \left\{t : t + \sum_{i=1}^t d_i \geq n \right\}.
\end{equation*}
\end{defn}

\begin{thm}[{\normalfont\cite{Slater}}]\label{SlaterThm}
For any graph $G$, $\gamma(G)\ge \sub(G)$, and this bound is sharp.
\end{thm}

For any $k\ge 1$, the $k$-domination number is monotonically increasing with respect to $k$; that is, $\gamma_k(G)\le \gamma_{k+1}(G)$. Keeping monotonicity in mind, it is natural that a parameter generalizing $\sub(G)$ will need to increase with respect to increasing $k$. This idea motivates the following definition. 

\begin{defn}\label{SubkDefn}
Let $k\ge 1$ be an integer, and $G$ be a graph. The \emph{sub-$k$-domination number} of $G$ is defined as
\begin{equation*}
\sub_k(G) = \min \left\{ t : t + \frac{1}{k}\sum_{i=1}^t d_i \geq n \right\}
\end{equation*}
\end{defn}

Since the vertex degrees of $G$ are integers between 0 and $n-1$, the sorted degree sequence of $G$ can be obtained in $O(n)$ time by counting sort (assuming vertex degrees can be accessed in $O(1)$ time). By maintaining the sum of the first $t$ elements in $D(G)$ and incrementing $t$, $\sub_k(G)$ can be computed in linear time; we state this formally below.

\begin{ob}
For any graph $G$ and positive integer $k$, $\sub_k(G)$ can be computed in $O(n)$ time.
\end{ob}

Taking $k=1$ in Definition \ref{SubkDefn},  we observe $\sub_1(G) = \sub(G)$, and hence $\sub_1(G)\le \gamma_1(G)$ by Theorem \ref{SlaterThm}. More generally, we will now  show that the $k$-domination number of a graph is bounded below by its sub-$k$-domination number.

\begin{thm} \label{sub-domination}
For any graph $G$ and positive integer $k$, $\gamma_k(G)\ge \sub_k(G)$, and this bound is sharp.
\end{thm}

\proof
Let $S = \{v_1,\ldots,v_t\}$ be a minimum $k$-dominating set of $G$. By definition, each of the $n-t$ vertices in $V(G)\backslash S$ is adjacent to at least $k$ vertices in $S$. Thus, the sum of the degrees of the vertices in $S$, i.e. $\sum_{i=1}^t d(v_i)$, is at least $k(n-t)$. Dividing by $k$ and rearranging, we obtain 
\begin{equation*}
t+\frac{1}{k}\sum_{i=1}^t d(v_i)\geq n.
\end{equation*}
Since the degree sequence of $G$ is non-increasing, it follows that $\sum_{i=1}^t d_i \geq \sum_{i=1}^t d(v_i)$. Thus, 
\begin{equation}
\label{eq1}
t + \frac{1}{k}\sum_{i=1}^t d_i \geq n.
\end{equation}
Since $\sub_k(G)$ is the smallest index for which (\ref{eq1}) holds, we must have $\sub_k(G) \leq t = \gamma_k(G)$.

When $k=1$, note that $\sub(K_{n-1,1}) = 1 = \gamma(K_{n-1,1})$. When $k>1$, let $G$ be a complete bipartite graph with a perfect matching removed where each part of the vertex partition is of size $k+1$. Then $\sub_k(G)=\min \{t:t+\frac{1}{k}\sum_{i=1}^t k \geq n\}=k+1=\gamma_k(G)$. Thus, the bound is sharp for all $k$.
\qed
\vspace{9pt}

In the next section, we compute $\sub_k(G)$ for several families of graphs and investigate graphs for which $\sub_k(G) = \gamma_k(G)$.

%%%%%%%%%%%%%%%%%%%%%%
\section{Graphs for which $\sub_k(G) = \gamma_k(G)$}
In this section we explore the case of equality for Theorem \ref{sub-domination}. First, note that $\sub(G) = \gamma(G) = n$ for an empty graph $G$. We therefore exclude empty graphs from the following discussion; that is, assume $\Delta\geq 1$. We begin with two observations for the case $k=1$.
\begin{prop} \label{equality max degree}
Let $G$ be a graph with $\Delta \geq n - 2$. Then, $\sub(G) = \gamma(G)$.
\end{prop}
\proof
If $\Delta = n-1$ then $\gamma(G) = 1$ and thus $\sub(G) = \gamma(G)$, since by Theorem \ref{sub-domination}, $1 \leq \sub(G) \leq \gamma(G) = 1$. If $\Delta = n-2$, then $\gamma(G) = 2$ since no single vertex can dominate the graph, but a maximum degree vertex and its non-neighbor is a dominating set. Moreover, $\sub(G)\neq 1$ since $1 + (n-2) < n$; thus, $2 \leq \sub(G) \leq \gamma(G) = 2$.
\qed
\vspace{9pt}

%If $\Delta = n-3$, $\gamma(G) \leq 3$ since any set containing a maximum degree vertex and its two non-neighbors is a dominating set for $G$. Moreover, if $\Delta \geq 1$, then $\sub(G) = 2$. To see this, first note that $2 \leq \sub(G)$ since $1 + n - 3 < n$. If $d_1 \geq d_2 \geq \cdots \geq d_n$ is the degree sequence of $G$, then suppose $2 + d_1 + d_2 < n$. By assumption, $d_1 = n - 3$, so we must have $d_2 < 1$, which is impossible. 

If $G$ is a graph with $\Delta \leq n-3$, then $\sub(G)$ may not be equal to $\gamma(G)$. For example, let $G$ be the graph obtained by appending a degree one vertex to two leaves of $K_{1,3}$; it can be verified that $\gamma(G)=3$ and $\sub(G)=2$.

\begin{prop} \label{equality domination at most 2}
Let $G$ be a graph with $\gamma(G) \leq 2$. Then $\sub(G) = \gamma(G)$.
\end{prop}
\proof
From Theorem \ref{sub-domination}, if $\gamma(G)=1$ then $\sub(G)=1$. Conversely, if $\sub(G)=1$, then $1+d_1\geq n$ and hence from Proposition \ref{equality max degree}, $\gamma(G)=1$. Similarly, if $\gamma(G)=2$ then $\sub(G)\leq 2$; however, since $\sub(G)=1$ if and only if $\gamma(G)=1$, it follows that $\sub(G)=2$.
\qed
\vspace{9pt}

If $G$ is a graph with $\gamma(G)\geq 3$, then $\sub(G)$ may not be equal to $\gamma(G)$. For example, let $G$ be the graph obtained by appending two pendants to each vertex of $K_3$; it can be verified that $\gamma(G)=3$ and $\sub(G)=2$.

We next characterize the sub-$k$-domination number of regular graphs. This will reveal some families of graphs for which $\sub_k(G) = \gamma_k(G)$ for $k \geq 2$.

\begin{thm} \label{regular sub-domination}
If $G$ is an $r$-regular graph, then $\sub_k(G) = \lceil \frac{kn}{r + k} \rceil$.
\end{thm}
\proof
Since $G$ is $r$-regular, $d_i = r$ for $1\leq i\leq n$. Then, from the definition of sub-$k$-domination, we have
\begin{equation}\label{def}
\sub_k(G) + \frac{\sub_k(G) r}{k}=\sub_k(G) + \frac{1}{k}\sum_{i=1}^{\sub_k(G)} d_i \ge n.
\end{equation}
Rearranging (\ref{def}), we obtain
\begin{equation}\label{reg2}
\frac{kn}{r + k} \leq \sub_k(G).
\end{equation}
Since $\sub_k(G)$ is the smallest integer that satisfies (\ref{reg2}), it follows that $\sub_k(G) = \lceil \frac{kn}{r + k} \rceil$. 
\qed
\vspace{9pt}

Note that $\gamma_k(G) = n$ whenever $k >\Delta(G)$. 
%However, as Theorem \ref{regular sub-domination} indicates, $\sub_k(G)$ may be less than $n$ even when $k$ is significantly larger than $\Delta$. 
We therefore restrict ourselves to the more interesting case of $k\leq \Delta$. The next example shows an infinite family of graphs for which the sub-$k$-domination number equals the $k$-domination number for all $k \leq \Delta$.

\begin{ob}
Let $C_n$ be a cycle. For all $k \leq \Delta$, $\sub_k(C_n)=\gamma_k(C_n)$.
\end{ob}
\proof
When $k=1$, it is known that $\gamma(C_n) = \big \lceil \frac{n}{3}\big \rceil$. Since cycles are $2$-regular, Theorem \ref{regular sub-domination} gives $\sub(C_n) = \big \lceil \frac{n}{3}\big \rceil$. Hence, $\gamma(C_n) = \sub(C_n)$ for all $n$. When $k=2$, Theorem \ref{regular sub-domination} gives $\lceil \frac{n}{2} \rceil \leq \sub_2(C_n)$. Since we can produce a 2-dominating set for $C_n$ by first picking any vertex $v$ and adding all vertices whose distance from $v$ is even, it follows that $\gamma_2(C_n) \leq \lceil \frac{n}{2} \rceil$. Thus $\sub_2(C_n) = \gamma_2(C_n)$. 
\qed
\vspace{9pt}

As another example, from Proposition \ref{equality max degree} and Theorem \ref{regular sub-domination}, we see that $\gamma(K_n) = \sub(K_n)=1$ and $\gamma_2(K_n)= \sub_2(K_n)=2$ for all $n$. When $k\geq 3$, $\gamma_k(K_n)$ does not equal $\sub_k(K_n)$ for all $n$ (for example, $\sub_3(K_4) = 2$ but $\gamma_3(K_4) = 3$); however, our next result shows that equality does hold when $n$ is large enough.

\begin{prop}\label{KnFail}
Let $K_n$ be a complete graph and let $k \leq n-1$ be a positive integer. Then $\sub_k(K_n) = \gamma_k(K_n)=k$ if and only if $n > (k-1)^2$.
\end{prop}
\proof 
First, note that $\gamma_k(K_n) = k$ for $k \leq n-1$, since any set of $k$ vertices of $K_n$ is $k$-dominating, while any set with at most $k-1$ vertices is at most $(k-1)$-dominating. Next, since $K_n$ is regular of degree $n-1$ it follows from Theorem \ref{regular sub-domination} that
\begin{equation*}
\sub_k(K_n) = \Big \lceil \frac{kn}{n-1+k} \Big \rceil \leq k = \gamma_k(K_n).
\end{equation*}
If $\sub_k(K_n) = k$, we must have
\begin{equation*}
\frac{kn}{n-1+k} > k - 1.
\end{equation*}
Rearranging, we obtain that $n > (k-1)^2$. \qed
\vspace{9pt}

Our last focus in this section is on the sub-$k$-domination number and $k$-domination number of $3$-regular, or \emph{cubic}, graphs. First, we recall an upper bound for the $k$-domination number due to Caro and Roditty \cite{CaroRoditty}.
\begin{thm}[\cite{CaroRoditty}]
Let $G$ be a graph, and $k$ and $r$ be positive integers such that $\delta \geq \frac{r+1}{r}k - 1$. Then, $\gamma_k(G) \leq \frac{r}{r+1}n$.
\end{thm}
\noindent In particular, for cubic graphs, Theorem \ref{regular sub-domination} and the Caro-Roditty bound (with $r$ taken to be the smallest positive integer satisfying $3\geq \frac{r+1}{r}k-1$) imply the following intervals for the $k$-domination number.
\begin{cor} \label{cubic interval}
Let $G$ be a cubic graph. Then,
\begin{enumerate}
\item $\lceil \frac{n}{4} \rceil \leq \gamma(G) \leq \lfloor \frac{n}{2} \rfloor$,
\item $\lceil \frac{2n}{5} \rceil \leq \gamma_2(G) \leq \lfloor \frac{n}{2} \rfloor$,
\item $\lceil \frac{n}{2} \rceil \leq \gamma_3(G) \leq \lfloor \frac{3n}{4} \rfloor$.
\end{enumerate}
\end{cor}
We see from Corollary \ref{cubic interval} that $\sub_k(G) = \gamma_k(G)$ for some cubic graphs with small values of $n$; for example, $\sub(G) = \gamma(G)$ when $n\leq 6$ and $\sub_2(G) = \gamma_2(G)$ when $n\leq 8$.

\section{Comparison to known bounds on $\gamma_k(G)$}
A well-known lower bound on the domination number of a graph is $\frac{n}{\Delta + 1}$. This bound is not difficult to derive \emph{a priori}, but it immediately follows from the definition of $\sub(G)$ and Theorem \ref{sub-domination}. In \cite{FinkJacobson}, Fink and Jacobson generalized this bound by showing that $\frac{kn}{\Delta + k} \leq \gamma_k(G)$; this also follows from a result of Hansberg and Pepper in \cite{Hansberg-Pepper}. In the following theorem, we show that $\sub_k(G)$ is an improvement on this bound.

\begin{thm} \label{n over max degree + 1}
Let $G$ be a graph; for every positive integer $k \leq \Delta$,
\begin{equation}
\label{lb1}
\frac{kn}{\Delta + k} \leq \sub_k(G) \leq \gamma_k(G).
\end{equation}
\end{thm}

\proof
The second inequality in (\ref{lb1}) follows from Theorem \ref{sub-domination}. To prove the first inequality, fix $k$ and let $t = \sub_k(G)$. By definition, $t + \frac{1}{k}\sum_{i=1}^t d_i \geq n$. Since $\Delta \geq d_i$ for $1\leq i\leq n$, it follows that 
\begin{equation*}
t + \frac{t\Delta}{k} =t + \frac{1}{k}\sum_{i=1}^t \Delta \geq t + \frac{1}{k}\sum_{i=1}^t d_i \geq n.
\end{equation*}
Rearranging the above inequality gives
\begin{equation*}
\frac{kn}{\Delta + k} \leq t = \sub_k(G).
\end{equation*}
\qed
\vspace{9pt}

Recall from Theorem \ref{regular sub-domination} that if $G$ is regular of degree $r$, then $\sub_k(G) = \lceil \frac{kn}{r + k} \rceil$. Thus, from Theorem \ref{n over max degree + 1}, we see that regular graphs minimize the sub-$k$-domination number over all graphs with $n$ vertices and maximum degree $\Delta$. This suggests that in order to maximize the sub-$k$-domination number, we might consider graphs which are, in some sense, highly irregular with respect to vertex degrees. This motivates the following theorem and its corollary.

\begin{thm} \label{sub domination glb}
Let $G$ be a graph; for $1 \leq t \leq \Delta$ let $n_t$ be the number of vertices of $G$ with degree $t$, let $s_t = \sum_{i=1}^t n_{\Delta + 1 - i}$, and let $\Delta_t = d_{s_t + 1}$. If $s_t + \sum_{i=1}^{s_t} d_i < n$ for some $t$, then
\begin{equation*}
\frac{kn - \sum_{i=1}^t(\Delta + 1 - \Delta_t - i)n_{\Delta + 1 - i}}{k + \Delta_t} \leq \sub_k(G).
\end{equation*}
\end{thm}

\proof
From the definition of $\sub_k(G)$, we have
\begin{equation} \label{glb ineq 1}
n\leq \sub_k(G) + \frac{1}{k} \sum_{i=1}^{\sub_k(G)} d_i.
\end{equation}
Since $s_t + \sum_{i=1}^{s_t} d_i < n$, it follows that $s_t < \sub_1(G)\leq \sub_k(G)$, and thus 
\begin{equation} \label{glb eq 1}
\sum_{i=1}^{\sub_k(G)} d_i = \sum_{i=1}^{s_t} d_i + \sum_{i = s_t+1}^{\sub_k(G)} d_i.
\end{equation}
Since $s_{t}=n_{\Delta}+n_{\Delta-1}+\ldots+n_{\Delta-t+1}$ and since the degree sequence of $G$ is non-increasing and has $n_j$ elements with value $j$, we have
\begin{eqnarray} 
\nonumber \sum_{i=1}^{s_t} d_i &=& \Delta n_{\Delta} + (\Delta - 1)n_{\Delta - 1} + \cdots + (\Delta - t + 1)n_{\Delta - t + 1} \\
&=& \sum_{i=1}^t (\Delta + 1 - i)n_{\Delta + 1 - i}. \label{glb eq 2}
\end{eqnarray}
Again since $D(G)$ is non-decreasing, we have that $\Delta_t = d_{s_t + 1}\geq d_{s_t + 2}\geq\cdots\geq d_{\sub_k(G)}$. Thus, it follows that
\begin{equation} \label{glb eq 3}
\sum_{i=s_t + 1}^{\sub_k(G)} d_i \leq \sum_{i=s_t + 1}^{\sub_k(G)} \Delta_t =(\sub_k(G) - s_t)\Delta_t.
\end{equation}
Substituting \eqref{glb eq 1}, \eqref{glb eq 2}, and \eqref{glb eq 3} into the right-hand-side of \eqref{glb ineq 1} yields
\begin{equation*}
n\leq \sub_k(G) + \frac{1}{k}\sum_{i=1}^t (\Delta + 1 - i)n_{\Delta + 1 - i} + \frac{1}{k} (\sub_k(G) - s_t)\Delta_t.
\end{equation*}
By expanding $(\sub_k(G) - s_t)\Delta_t$ and substituting $s_t = \sum_{i=1}^t n_{\Delta + 1 - i}$, the above inequality can be rewritten as
\begin{equation*}
n\leq \sub_k(G) \Big ( 1 + \frac{\Delta_t}{k} \Big ) + \frac{1}{k} \sum_{i=1}^t (\Delta + 1 - \Delta_t - i)n_{\Delta + 1 -i }.
\end{equation*}
Rearranging the preceding inequality gives
\begin{equation*}
\frac{kn - \sum_{i=1}^t (\Delta + 1 - \Delta_t - i)n_{\Delta + 1 - i}}{k + \Delta_t} \leq \sub_k(G).
\end{equation*}
\qed
\vspace{9pt}

We note that the bound in Theorem \ref{sub domination glb} is optimal when $t$ is taken to be the maximum positive integer for which $s_t + \sum_{i=1}^{s_t} d_i < n$. Theorem \ref{sub domination glb} can be used to give simple lower bounds for the $k$-domination number of a graph when certain restrictions on the order and maximum degree are met. These bounds also improve on the lower bound given in Theorem \ref{n over max degree + 1}.

\begin{cor} \label{sub-domination not enough max degree}
Let $G$ be a graph, let $n_\Delta$ denote the number of maximum degree vertices of $G$, and let $\Delta'$ denote the second-largest degree of $G$. If $k$ is a positive integer and $n_{\Delta} + \frac{\Delta n_{\Delta}}{k} < n$, then
\begin{equation}
\label{eq_cor_lb}
\frac{kn - n_{\Delta}(\Delta - \Delta')}{\Delta' + k} \leq \sub_k(G) \leq \gamma_k(G).
\end{equation}
\end{cor}

\proof
Take $t = 1$ in the bound from Theorem \ref{sub domination glb} and note that $s_1=n_\Delta$ and $\Delta_1=d_{n_\Delta+1}=\Delta'$. Since $n_{\Delta} + \frac{\Delta n_{\Delta}}{k} < n$, we have that $s_1 + \frac{1}{k} \sum_{i=1}^{s_1} d_i =n_{\Delta} + \frac{1}{k} \sum_{i=1}^{n_{\Delta}} d_i=n_\Delta+\frac{\Delta n_\Delta}{k}< n$. Thus, the condition of Theorem \ref{sub domination glb} is satisfied, and we obtain the first inequality in (\ref{eq_cor_lb}); the second inequality in (\ref{eq_cor_lb}) follows from Theorem \ref{sub-domination}. 
\qed
\vspace{9pt}

We see from Corollary \ref{sub-domination not enough max degree} that if $G$ has a unique maximum degree vertex, then
\begin{equation*}
\frac{kn - \Delta + \Delta'}{\Delta' + k} \leq \gamma_k(G).
\end{equation*}
Corollary \ref{sub-domination not enough max degree} gives significant improvements on the lower bound in Theorem \ref{n over max degree + 1} whenever the difference between $\Delta$ and $\Delta'$ is large. For example, consider the corona of $K_{1,n-1}$ ($n \geq 3$) which is obtained by appending a vertex of degree 1 to each of the $n-1$ vertices of degree 1 in $K_{1,n-1}$. The degree sequence of this graph is $ \{ n-1, 2^{n-1}, 1^{n-1} \} $ and its order is $2n-1$. This graph meets the conditions of Corollary \ref{sub-domination not enough max degree}, and the bound given in the corollary simplifies to $\frac{(2k-1)n - (k-3)}{2 + k}$, whereas the bound given by Theorem \ref{n over max degree + 1} is $\frac{k(2n-1)}{n - 1 + k}$. To compare these two bounds, we first compute the difference between them:
\begin{equation*}
\frac{(2k-1)n-(k-2)}{2+k} - \frac{k(2n-1)}{n-1+k} = \frac{(2k-1)n^2 + (4-6k)n + 8k - k^2 - 3}{(2+k)(n-1+k)}.
\end{equation*}
When $k$ is fixed, the difference between these two bounds approaches $\infty$ as $n \to \infty$.

\section{Critical graphs}

There are three natural ways to consider critical graphs in the context of sub-$k$-domination: graphs which are critical with respect to edge-deletion, edge-addition, and vertex-deletion. 

\begin{defn}
Let $G$ be a graph and $k$ be a positive integer. We will say that 
\begin{enumerate}
\item $G$ is \emph{edge-deletion-$\sub_k(G)$-critical} if for any $e\in E(G)$, $\sub_k(G-e)> \sub_k(G)$.
\item $G$ is \emph{edge-addition-$\sub_k(G)$-critical} if for any $e\in E(\overline{G})$, $\sub_k(G+e)<\sub_k(G)$. 
\item $G$ is \emph{vertex-deletion-$\sub_k(G)$-critical} if for any $v\in V(G)$, $\sub_k(G-v)>\sub_k(G)$. 
\end{enumerate}
These properties will respectively be abbreviated as \emph{$\sub_k(G)$-ED-critical}, \emph{$\sub_k(G)$-EA-critical}, and \emph{$\sub_k(G)$-VD-critical}.
\end{defn}

In this section, we present several structural results about sub-$k$-domination critical graphs, including connections to other graph parameters. Throughout the section, we will assume that given a graph $G$ with $V(G)=\{v_1,\ldots,v_n\}$ and $D(G)=\{d_1,\ldots, d_n\}$ where $d_1\geq\cdots\geq d_n$, it holds that $d_i=d(v_i)$ --- in other words, the vertices of $G$ are labeled according to a non-increasing ordering of their degrees.

We first present two results about $\sub_k(G)$-ED-critical graphs.

\begin{prop}
\label{indep set prop}
Let $G$ be a $\sub_k(G)$-ED-critical graph with $\sub_k(G)=t$. Then $\{v_{t+1},\ldots,v_n\}$ is an independent set of $G$, and $n-\sub_k(G)\leq \alpha(G)$. 
\end{prop}

\proof
Suppose for contradiction that $\{v_{t+1},\ldots,v_n\}$ is not an independent set and let $e=v_xv_y$ be an edge with $v_x,v_y\in\{v_{t+1},\ldots,v_n\}$. Then, the degree sequence of $G-e$ is $d_1'\geq \cdots\geq d_n'$, where $d_i'=d_i$ for all $1\leq i\leq t$. Thus, $t+\frac{1}{k}\sum_{i=1}^t d_i'=t+\frac{1}{k}\sum_{i=1}^t d_i\geq n$, which implies that $\sub_k(G-e)\leq t$; this contradicts the assumption that $G$ is $\sub_k(G)$-ED-critical. Thus, $\{v_{t+1},\ldots,v_n\}$ is an independent set, so $\alpha(G)\geq n-t$.
\qed
\vspace{9pt}

\begin{prop}
Let $G$ be a $\sub_k(G)$-ED-critical graph with no isolates and $\sub_k(G)=t$. Then $\lfloor t+\frac{1}{k}\sum_{i=1}^td_i\rfloor=n$, and for any $e\in E(G)$, $\sub_k(G-e)=\sub_k(G)+1$. 
\end{prop}

\proof
By definition of $\sub_k(G)$ and since $n$ is an integer, we have that $\lfloor t+\frac{1}{k}\sum_{i=1}^td_i\rfloor\geq n$. Suppose for contradiction that $\lfloor t+\frac{1}{k}\sum_{i=1}^td_i\rfloor> n$. Since by Proposition \ref{indep set prop}, $\{v_{t+1},\ldots,v_n\}$ is an independent set of $G$ and since $G$ has no isolates, we can choose an edge $e$ incident to exactly one vertex in $\{v_1,\ldots,v_t\}$. The degree sequence of $G-e$ is $d_1'\geq \cdots\geq d_n'$, where $\sum_{i=1}^t d_i'=(\sum_{i=1}^t d_i)-1$. Thus, 

\begin{equation*}
t+\frac{1}{k}\sum_{i=1}^t d_i'=t+\frac{1}{k}\sum_{i=1}^t d_i-\frac{1}{k}\geq \left\lfloor t+\frac{1}{k}\sum_{i=1}^t d_i\right\rfloor-\frac{1}{k}\geq n+1-\frac{1}{k}\geq n,
\end{equation*}
meaning $\sub_k(G-e)=t$, which contradicts $G$ being $\sub_k(G)$-ED-critical.

Now let $e$ be any edge of $G$ and $d_1'\geq \cdots\geq d_n'$ be the degree sequence of $G-e$. The deletion of $e$ decreases $\sum_{i=1}^{t+1} d_i$ by at most $2$, i.e., $\sum_{i=1}^{t+1} d_i'\geq(\sum_{i=1}^{t+1} d_i)-2$. Thus,

\begin{equation*}
(t+1)+\frac{1}{k}\sum_{i=1}^{t+1}d_i'\geq (t+1)+\frac{1}{k}\sum_{i=1}^{t+1}d_i-\frac{2}{k}= t+\frac{1}{k}\left(\sum_{i=1}^td_i\right)+\frac{d_{t+1}-2}{k}+1\geq n,
\end{equation*}
where in the last inequality $d_{t+1}\geq 1$ since $G$ has no isolates; this implies $\sub_k(G-e)=t+1=\sub_k(G)+1$.
\qed
\vspace{9pt}

Next, we present two analogous results about $\sub_k(G)$-EA-critical graphs.

\begin{prop}\label{completeSameDegree}
Let $G$ be a $\sub_k(G)$-EA-critical graph with $\sub_k(G)=t$. Then the vertices in $\{v\in V(G):d(v)<d_t\}$ form a clique. 
\end{prop}

\proof
Suppose on the contrary that there are two non-adjacent vertices $v_x$ and $v_y$ with $d_t>d_x\geq d_y$. Then, the degree sequence of $G+v_xv_y$ is $d_1'\geq \cdots\geq d_n'$, where $d_i'=d_i$ for all $1\leq i\leq t$. This implies that $\sub_k(G+e)=\sub_k(G)$, a contradiction. 
\qed
\vspace{9pt}

\begin{prop}
Let $G$ be a $\sub_k(G)$-EA-critical graph with no isolates and $\sub_k(G)=t$. Then, for each $e\in E(\overline{G})$, $\sub_k(G+e)=\sub_k(G)-1$.
\end{prop}

\proof
Let $e$ be any edge in $\overline{G}$ and $d_1'\geq \cdots\geq d_n'$ be the degree sequence of $G+e$. The addition of $e$ increases $\sum_{i=1}^{t} d_i$ by at most $2$, i.e., $\sum_{i=1}^{t} d_i'\leq(\sum_{i=1}^{t} d_i)+2$. Thus, 

\begin{equation*}
(t-2)+\frac{1}{k}\sum_{i=1}^{t-2} d_i'=(t-2)+\frac{1}{k}\sum_{i=1}^t d_i'-\frac{d_t'+d_{t-1}'}{k}\leq t+\frac{1}{k}\sum_{i=1}^t d_i-\frac{d_t'+d_{t-1}'}{k}\leq n-\frac{d_t'+d_{t-1}'}{k}<n,
\end{equation*}
where in the last inequality $\frac{d_t'+d_{t-1}'}{k}>0$ since $G$ has no isolates; this implies $\sub_k(G+e)>t-2$, so $\sub_k(G+e)=\sub_k(G)-1$.
\qed
\vspace{9pt}

Graphs that are $\sub_k(G)$-VD-critical differ from $\sub_k(G)$-ED-critical graphs and $\sub_k(G)$-EA-critical graphs, in the sense that it is possible for $\sub_k(G-v)$ and $\sub_k(G)$ to differ by much more than 1. For example, this is the case for the star $K_{n-1,1}$ when the center of the star is the vertex removed. We now show another result for $\sub_k(G)$-VD-critical graphs.

%\begin{prop}
%Let $G$ be a $\sub_k(G)$-VD-critical graph with $\sub_k(G)=t$. Then there is at least one edge joining a vertex in $\{v_{t+1},\ldots,v_n\}$ to a vertex in $\{v_1,\ldots,v_t\}$.
%\end{prop}
%
%\proof
%Suppose that $v_x\in\{v_{t+1},\ldots,v_n\}$ is not adjacent to any vertices in $\{v_1,\ldots,v_t\}$. Then $G-v_x$ has degree sequence $d_1'\geq\ldots\geq d_{n-1}'$ where $d_i'=d_i$ for $1\leq i\leq t$. Thus, $t+\frac{1}{k}\sum_{i=1}^t d_i'=t+\frac{1}{k}\sum_{i=1}^t d_i\geq n\geq n-1$, which implies that $\sub_k(G-v_x)\leq t$; this contradicts the assumption that $G$ is $\sub_k(G)$-VD-critical.
%\qed
%\vspace{9pt}

%\begin{prop}
%Let $G$ be a $\sub_k(G)$-VD-critical graph with $\sub_k(G)=t$ and let $x\in\{1,\ldots,t-1\}$. If $d_x>d_t$ then $\left(t+\frac{1}{k}\sum_{i=1}^td_i\right)-d_x<n$. If $d_x=d_t$ and $d_x$ is adjacent to some vertex in $\{d_1,\ldots,d_t\}\setminus\{d_x\}$ then $\left(t+\frac{1}{k}\sum_{i=1}^td_i\right)-d_x<n$. 
%\end{prop}

\begin{prop}
Let $G$ be a $\sub_k(G)$-VD-critical graph with $\sub_k(G)=t$. Then each vertex in $\{v_{t+1},\ldots,v_n\}$ is adjacent to at least $k+1$ vertices in $\{v_1,\ldots,v_t\}$.
\end{prop}

\proof
Suppose that $v_x\in\{v_{t+1},\ldots,v_n\}$ is adjacent to at most $k$ vertices in $\{v_1,\ldots,v_t\}$. Then $G-v_x$ has degree sequence $d_1',\ldots,d_{n-1}'$ such that $\sum_{i=1}^t d_i'\geq (\sum_{i=1}^t d_i)-k$. Thus, $t+\frac{1}{k}\sum_{i=1}^t d_i'\geq t+\frac{1}{k}\sum_{i=1}^t d_i-1\geq n-1$, which implies that $\sub_k(G-v_x)\leq t$; this contradicts the assumption that $G$ is $\sub_k(G)$-VD-critical.
\qed

%%%%%%%%%%
\section{Conclusion}
In this paper, we introduced the sub-$k$-domination number and showed that it is a computationally efficient lower bound on the $k$-domination number of a graph. We also showed that the sub-$k$-domination number improves on several known bounds for the $k$-domination number, and gave some conditions which assure that $\sub_k(G) = \gamma_k(G)$. This investigation was a step toward the following general question:

\begin{prob}
For each positive integer $k$, characterize all graphs for which $\gamma_k(G)=\sub_k(G)$. 
\end{prob}

As another direction for future work, it would be interesting to define and study an analogue of sub-$k$-domination which is an upper bound to the $k$-domination number, or explore degree sequence based invariants which bound the connected domination number or the independent domination number of a graph.

\section*{Acknowledgements}
This work is supported by the National Science Foundation, Grant No. 1450681 (B. Brimkov).


\medskip
\begin{thebibliography}{99}

\bibitem{DSI} 
Y. Caro and R. Pepper, 
Degree sequence index strategy, 
\emph{Australasian Journal of Combinatorics}, 
59: 1--23 (2014). 

\bibitem{CaroRoditty}
Y. Caro and Y. Roditty, 
A note on the $k$-domination number of a graph,
\emph{International Journal of Mathematics and Mathematical Sciences},
13: 205--206 (1990).

\bibitem{TotalIntro} 
E.J. Cockayne, R.M. Dawes, and S.T. Hedetniemi, 
Total domination in graphs,
\emph{Networks}, 
10: 211--219 (1980).

\bibitem{Delavina} 
E. DeLaVi\~{n}a, C.E. Larson, R. Pepper, and B. Waller, 
Graffiti.pc on the $2$-domination number of a graph,
\emph{Congressus Numerantium},
203: 15--32 (2010).

\bibitem{Dorfling}
M. Dorfling, W.D. Goddard, and M. Henning, 
Domination in planar graphs with small diameter II,
\emph{Ars Combinatoria}, 
78: 237--255 (2006).


\bibitem{Du}
Z. Du, and A. Ili\'{c}, 
A proof of the conjecture regarding the sum of domination number and average eccentricity,
\emph{Discrete Applied Mathematics}, 
201: 105--113 (2016).


\bibitem{FavaronResidue} 
O. Favaron, M. Mah\v{e}o, and J.F. Sacl\v{e}, 
On the residue of a graph,
\textit{Journal of Graph Theory},
15: 39--64 (1991).


\bibitem{Favaron} 
O. Favaron, A. Hansberg, and L. Volkmann, 
On the $k$-domination and minimum degree in graphs,
\emph{Journal of Graph Theory},
57: 33--40 (2008).
    
\bibitem{FinkJacobson} 
J.F. Fink and M.S. Jacobson, 
$n$-Domination in graphs,
\emph{Graph Theory with Applications to Algorithms and Computer Science (Kalamazoo, Mich., 1987)},
pp. 283--300, Wiley, 1985.    


\bibitem{Glebov}
R. Glebov, A. Liebenau, and T. Szab\'{o}, 
On the concentration of the domination number of the random graph,
\emph{SIAM Journal on Discrete Mathematics}, 
29(3): 1186--1206 (2015).


\bibitem{DomBook1} 
T. Haynes, S.T. Hedetniemi and P.J. Slater, 
Fundamentals of Domination in Graphs. Marcel Decker, Inc., NY, 1998.

\bibitem{DomBook2} 
T. Haynes, S.T. Hedetniemi and P.J. Slater, 
Domination in Graphs: Advanced Topics. Marcel Decker, Inc., NY, 1998.
 
\bibitem{Adriana} 
A. Hansberg, 
Bounds on the connected $k$-domination number in graphs,
\emph{Discrete Applied Mathematics},
158(14): 1506--1510 (2010).

\bibitem{Hansberg-Pepper} 
A. Hansberg and R. Pepper, 
On $k$-domination and $j$-independence in graphs,
\emph{Discrete Applied Mathematics},
161: 1472--1480 (2013).
 
\bibitem{MHAYbookTD} 
M. A. Henning and A. Yeo, 
\emph{Total domination in graphs (Springer Monographs in Mathematics)}.  ISBN-13: 978-1461465249 (2013).

\bibitem{Isaacs} 
R. Isaacs, 
Infinite families of nontrivial trivalent graphs which are not Tait colorable,
\emph{American Mathematical Monthly},
82: 221--239 (1975).
 
\bibitem{NP} 
M.S. Jacobson and K. Peters, 
Complexity questions for $n$-domination and related parameters,
\emph{Congressus Numerantium},
68: 7--22 (1989). 
 
\bibitem{Jelen} 
F. Jelen, 
$k$-independence and the $k$-residue of a graph,
\textit{Journal of Graph Theory},
127: 209--212 (1999).
 
\bibitem{Mekis} 
G. Meki\v{s}, 
Lower bounds for the domination number and the total domination number of direct product graphs.  
\textit{Discrete Mathematics}, 
310: 3310--3317 (2010).
 
\bibitem{PepperThesis} 
R. Pepper, 
Binding independence.
\textit{PhD Thesis}
University of Houston (2004).

 
\bibitem{Pepper} 
R. Pepper, 
Implications of some observations about the $k$-domination number,
\emph{Congressus Numerantium},
206: 65--71 (2010).
 
\bibitem{Rautenbach} 
D. Rautenbach and L. Volkmar, 
New bounds on the $k$-domination number and the $k$-tuple domination number, 
\emph{Applied Mathematics Letters},
20: 98--102 (2007).
 
\bibitem{Slater} 
P.J. Slater, 
Locating dominating sets and locating-dominating sets,
\emph{Graph Theory, Combinatorics, and Applications: Proceedings of the 7th Quadrennial International Conference on the Theory and Applications of Graphs},
2: 1073--1079 (1995).

\bibitem{Stevanovic}
D. Stevanovi\'{c}, M. Aouchiche, and P. Hansen, 
On the spectral radius of graphs with a given domination number,
\emph{Linear Algebra and its Applications},
428(8): 1854--1864 (2008).


\end{thebibliography}
\end{document}